# Paris Agreement requires substantial, broad, and sustained engagements beyond COVID-19 public stimulus packages

Katsumasa Tanaka[1,2], Christian Azar[3], Olivier Boucher[4],

Philippe Ciais[1], Yann Gaucher[1], and Daniel J. A. Johansson[3]

[1] Laboratoire des Sciences du Climat et de l'Environnement (LSCE), IPSL, CEA/CNRS/UVSQ, Université Paris-Saclay, Gif-sur-Yvette, France

[2] Earth System Risk Analysis Section, Earth System Division, National Institute for Environmental Studies (NIES), Tsukuba, Japan

[3] Division of Physical Resource Theory, Department of Space, Earth, and Environment, Chalmers University of Technology, Gothenburg, Sweden

[4] Institut Pierre-Simon Laplace (IPSL), CNRS / Sorbonne Université, Paris, France

Corresponding authors: Katsumasa Tanaka (katsumasa.tanaka@lsce.ipsl.fr) and Christian Azar (christian.azar@chalmers.se)

**Abstract**

It has been claimed that COVID-19 public stimulus packages could be sufficient to meet the short-term energy investment needs to leverage a shift toward a pathway consistent with the 1.5 °C target of the Paris Agreement. Here we provide complementary perspectives to reiterate that substantial, broad, and sustained engagements beyond stimulus packages will be needed for achieving the Paris Agreement long-term targets. Low-carbon investments will need to scale up and persist over the next several decades following short-term stimulus packages. The required total energy investments in the real world can be larger than the currently available estimates from Integrated Assessment Models (IAMs). Existing databases from IAMs are not sufficient for analyzing the effect of public spending on emission reduction. To inform what role COVID-19 stimulus packages and public investments may play for reaching the Paris Agreement targets, explicit modelling of such policies is required.

**Acknowledgments**

We are grateful to Christoph Bertram (PIK, Germany) for sharing with us the data used in Fig. 2 and providing useful comments on this manuscript.

**Declarations**

**Author contributions**: All authors contributed to the study conception and design. Katsumasa Tanaka and Christian Azar coordinated the study. The analysis was performed by Katsumasa Tanaka, Olivier Boucher, Yann Gaucher, and Daniel J. A. Johansson. All authors discussed the results and contributed to developing the argument. Katsumasa Tanaka drafted the manuscript, with input from all other authors. All authors read and approved the final manuscript.

**Funding**: K.T. benefited from State assistance managed by the National Research Agency in France under the Programme d'Investissements d'Avenir under the reference ANR-19-MPGA-0008. D.J.A.J. acknowledges Mistra for funding through the research program Mistra Carbon Exit. C.A. acknowledges financial support from MISTRA electric transition and Carl Bennet Foundation AB.

**Data availability:** Data supporting the conclusions are present in the paper. Data and Python code used to generate the figures in the paper, as well as the investment data shared by C. Bertram, are available on Zenodo with doi.org/10.5281/zenodo.5104289.

**Conflict of interest**: K.T. is an Associate Deputy Editor of *Climatic Change*. His status has no bearing on editorial consideration. The other authors declare no competing interests.

# 1 Introduction

In the current context, it appears quite natural to use a subset of COVID-19 public stimulus packages for green investments in order to steer the world towards sustainability. Integrated assessment models (IAMs) that combine economy, energy, climate, and sometimes also land-use models have been used to inform such debate. IAMs typically simulate how the long-term temperature goals of the Paris Agreement could be met, what type of investment fulfillment would entail, and how large the associated costs would be. These models seek to find the lowest social cost under a carbon price pathway that leads to the long-term goals.

Shortly after the start of the COVID-19 pandemic, some studies (Andrijevic et al. 2020; Climate Action Tracker 2020; Forster et al. 2020) discussed the potential of COVID-19 stimulus funds for promoting a transformation towards the 1.5°C target of the Paris Agreement on the basis of short-term energy investments simulated by such IAMs. Here we argue that energy investments in IAMs need to be interpreted with care by analyzing primarily Andrijevic et al. (2020) published in *Science* (thereafter, A20), as well as two more recent datasets from the Network for Greening the Financial System (NGFS) project (Bertram et al. 2021a) and the Exploring National and Global Actions to reduce Greenhouse gas Emissions (ENGAGE) project (Bertram et al. 2021b). More than a year later, this topic continues to be relevant as both the COVID-19 recovery and the transition onto a pathway compatible with the Paris Agreement goals are far from over (IEA 2021b).

A20 compared the COVID-19 public stimulus funds around the world (12.2 trillion US$ globally at the time of A20) with the estimates of necessary energy investments indicated by IAMs. They claimed an estimate of 300 billion US$/year until 2024 as the additional investments for low-carbon energy technologies and energy efficiency required globally in order to leverage a shift from a current pathway (reflecting stated policies until 2030) to an ambitious pathway aiming for the 1.5 °C target (thereafter, *additional low-carbon investments*). By taking into consideration the reduction of investments in fossil fuels, they further claimed an estimate of 20 billion US$/year until 2024 (thereafter, *additional total energy investments*). They concluded that "in sum, a small fraction of announced COVID-19 economic recovery packages could provide the necessary financial basis for a decided shift toward a Paris Agreement-compatible future." Although we agree with A20 and others that COVID-19 stimulus funds may offer an opportunity to boost climate actions (Hepburn et al. 2020) and may be important for reducing upfront risks that can deter low-carbon investments (Hourcade et al. 2021), we nevertheless believe that the conclusions by A20 misrepresent the grand challenges that climate change mitigation entails (IPCC 2018). In our view, their analysis and other similar claims need to be balanced by the following five arguments.

## 2 Five arguments

### 2.1 Need for accelerating low-carbon investments in decades to come

Stimulus packages are only short-term actions, while investments will need to scale up and persist over the next several decades to continue to develop low-carbon energy technologies and increase energy efficiency, among other transformation needs (IEA 2017; European Commission 2018; IPCC 2018). We confirm this point by analyzing the data by McCollum et al. (2018), which A20 rely on. In Fig. 1a, the mean projection of IAMs indicates a need for accelerating low-carbon investments in decades to come to follow a 1.5 °C target pathway. In fact, A20 presented in Fig. S8 and S9 that the additional low-carbon investments until 2050 would be on average four to five times larger than those until 2024 in annual terms. Despite this, they omitted to consider the long-term investment requirements when drawing their conclusions. To complement the argument of A20, we thus argue that accelerating low-carbon investments is also required in decades to come beyond the near term.

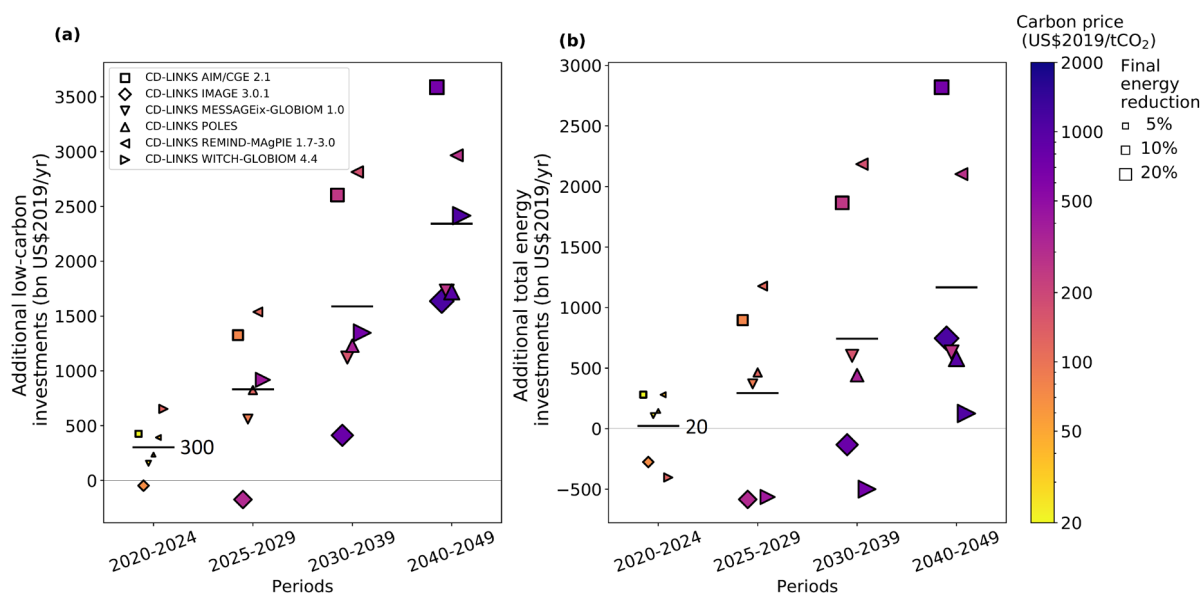

**Fig. 1 Additional low-carbon and total energy investments required for achieving the 1.5 °C warming target relative to current policy levels.** See text for the definitions of the *additional low-carbon investments* and *additional total energy investments*. Estimates obtained from individual IAMs are shown in symbols according to the legend; the model-means are in horizontal black bars. The estimates of 300 and 20 billion US$2019/year highlighted in A20 are indicated beside the respective black bars. The global carbon price (on a logarithmic scale) assumed in each IAM is presented according to the color scale. The marker size is proportional to the absolute percentage reduction in final energy demand (normalized to 10% reduction), relative to the level under the current pathway reflecting stated policies until 2030. Data were obtained from the CD-LINKS database (McCollum et al. (2018);

https://db1.ene.iiasa.ac.at/CDLINKSDB/; NPi2020_400 and NPi scenarios for 1.5 °C pathways (with high overshoot) and current pathways, respectively), aggregated over the four different periods through linear interpolation (e.g. the estimate for the period 2020-2024 is the sum of the data for year 2020 weighted by 0.6 and the data for year 2025 weighted by 0.4), and adjusted for inflation (a factor of 1.16 and 1.08 applied to update the unit from US$2010 and US$2015, respectively, to US$2019). Total energy investments comprise fossil fuel and low-carbon investments. The estimations of fossil fuel and low-carbon investments follow the respective definitions of A20: namely, fossil fuel investments account for "extraction and conversion of fossil fuels, electricity from fossil fuels without Carbon Capture and Storage (CCS) technologies and hydrogen from fossil fuels." Low-carbon investments consider "extraction and conversion of nuclear energy, CCS, electricity from non-bio renewables, hydrogen from non-fossil fuels, extraction and conversion of bioenergy, electricity transmission and distribution and storage, and energy efficiency."

## 2.2 High near-term carbon prices and their effects on energy demand and investments

The required additional total energy investments in the real world are highly uncertain but can be much larger than what A20 characterized. Fig. 1b indicates that the net 20 billion US$/year estimate is, according to our analysis, the mean of several larger values of opposing signs (between -400 and 280 billion US$/year). The amount of 20 billion US$/year corresponds, roughly speaking, merely to the costs of building a few nuclear power plants every year (Lovering et al. 2016). Such a surprisingly small mean value is influenced by two IAMs that assume very high global carbon prices (70 and 127 US$/tCO$_2$) already in the current period from 2020 to 2024. In reality, such high carbon prices have been achieved only in Europe very recently (i.e. EU emissions trading system (ETS) price of near 100 Euro/tCO$_2$ as of this writing in early February 2022). Only 22% of the greenhouse gas emissions around the world are currently covered by carbon pricing (World Bank 2021), giving an average price for global emissions of just 3 US$/tCO$_2$ (Parry et al. 2021). Carbon prices implemented explicitly or implicitly in the IAMs automatically incentivize low-carbon investments and disincentivize fossil fuel investments at the same time, but they also lower energy demand. Rapid reductions in final energy demand (i.e. data points indicated in large markers in Fig. 1b) due to the high near-term carbon prices might have led to the reductions in additional total energy investments until 2030s in the subset of IAMs. In the long run, however, additional total energy investments will increase also in the subset of IAMs due to growing investments on low-carbon sources, while energy demand will be lower than the level under the current pathway (Scott et al. 2022).

We argue here that such model results driven by high near-term carbon prices do not correspond to a realistic short-term pathway compatible with long-term requirements (van Vuuren et al. 2010). The reductions in additional total energy investments until 2030s in these IAMs are due to the assumed high near-term carbon prices that caused large reductions in final energy demand. In our view, such results are not suitable for informing the near-term investment needs toward the 1.5 °C target. Hence, we argue that models that can simulate 1.5 °C target pathways with near-term carbon prices consistent with those existing in the real world should be used for such a purpose. With only the subset of IAMs that used more moderate near-term carbon prices, the required additional total energy investments during 2020-2024 is estimated to be about 200 billion US$/year, substantially higher than the estimate of 20 billion US$/year based on all IAMs given by A20.

Two more recent datasets from NGFS and ENGAGE based on IAMs that include recent trends and understanding (Creutzig et al. 2017; Riahi et al. 2021) yield comparably higher estimates of the additional total energy investments for 2020-2024 (335 and 227 billion US$/year; green and black horizontal bars in Fig. 2b, respectively). On the other hand, our argument that very high near-term carbon prices suppress final energy demand and further reduce additional total energy investments does not explain some of the model results. The response of final energy demand to near-term carbon prices varies across models, with POLES in ENGAGE indicating little change in final energy demand under very high near-term carbon prices (i.e. 523 and 1,661 US$/tCO$_2$ in 2020-2024 and 2025-2029, respectively). In the case of WITCH in ENGAGE, oil investments plummet under the 1.5 °C pathway over decades, which might have led to the negative additional total energy investments throughout the period (C. Bertram, *personal communication*). This indicates that our argument holds only under certain model assumptions and scenarios. More in-depth studies are needed to investigate the relationship between the carbon price and the energy demand and investments for understanding the investments needed if emission reduction policies are primarily driven by public support rather than carbon pricing.

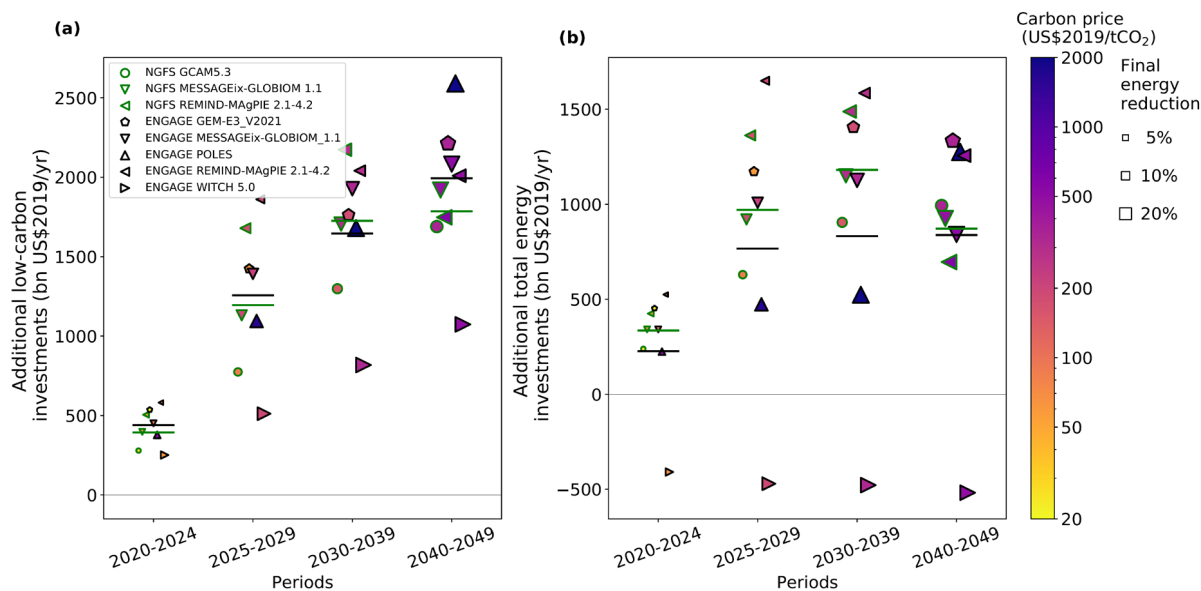

**Fig. 2 Results comparable with Fig. 1, based on more recent datasets from the NGFS and ENGAGE projects.** Regarding the results from NGFS, all data were taken from the NGFS database (https://data.ene.iiasa.ac.at/ngfs/; the Net Zero 2050 scenarios and the Current Policies scenarios interpreted as 1.5 °C pathways (without or with low overshoot) and current pathways, respectively, in our analysis). Low-carbon investments are those into the low-carbon energy supply system and energy efficiency of energy demand technologies. Total energy investments are those into the energy system including energy efficiency. Regarding the results from ENGAGE, data for carbon price and final energy demand were obtained from the ENGAGE database (https://data.ene.iiasa.ac.at/engage/; EN_NPi2020_600 and EN_NPi2100 scenarios interpreted as 1.5 °C pathways (without or with low overshoot) and current pathways, respectively, in our analysis). Data for low-carbon and total energy investments are obtained from C. Bertram. Only a subset of models from ENGAGE is shown due to data availability. Low-carbon investments include those related to electricity supply from wind, solar, and other low-carbon sources, and supply system (transportation, storage, and distribution), as well as those related to energy efficiency improvement. Fossil fuel investments include those related to electricity supply from fossil fuels, as well as those related to extraction of natural gas, oil, and coal. The sum of low-carbon and fossil fuel investments gives total energy investments. For both datasets, the same temporal aggregation and inflation correction factor with those in Fig. 1 were applied. Investment categories in the two groups are not exactly same with those in Fig. 1.

## 2.3 Carbon pricing as a backbone of decarbonization

Energy investments into low-carbon energy and away from fossil fuels are most cost-effectively induced by carbon pricing complemented with subsidies for technology development and the expansion of low-carbon infrastructure (Sandén and Azar 2005; Baranzini et al. 2017; Hourcade et al. 2021), unlike what is implicitly assumed when comparing the face values of recovery public funds with energy investments in IAMs. It is well-established that carbon pricing should be the backbone to meet the Paris Agreement targets cost-effectively (Stiglitz et al. 2017). Evidence suggests that it would be highly unwarranted to cover all energy investments by public funds (Kalkuhl et al. 2013). The International Energy Agency projects that more than 70% of clean energy and electricity network investments will come from private funds under its sustainable development scenario until 2030 (IEA 2021b).

## 2.4 Need for explicit modeling of public spending

In order to better inform what role COVID-19 stimulus packages and public investments may play for reaching the Paris Agreement temperature targets, models need to simulate such policies explicitly to analyze to what extent such policies would complement carbon pricing and how such policies would impact energy investments and energy prices. The IAMs used by McCollum et al. and A20 were driven by a carbon price pathway under a specific carbon budget (McCollum et al. 2018). One can expect that such carbon prices should generate significant private capital flows to support energy investments without requiring substantial public funds. This indicates an inconsistency between the policy framework presented and the investment estimate provided. Thus, if A20 is in favor of stringent carbon pricing (as in the models employed), then the need for public support packages towards decarbonization would be reduced since carbon pricing would provide incentives for private capital to carry out the required investments. If, instead, the proposed policy framework is one in which public support packages are driving the transition, then the estimated investment levels are likely too low (since they were based on models where carbon pricing reduced the energy demand and consequently necessary energy investments).

It should be noted that we do not argue that IAMs cannot be used to assess COVID-19 stimulus packages and public investments. Rather, we argue that mitigation pathways generated by public support packages may turn out to be very different from those generated by an explicit implementation of a carbon price pathway or those by optimization towards a carbon budget (or other type of climate targets) (Johansson et al. 2020). Public support-driven pathways may lead to a different energy system development with different energy investments, energy prices, and social costs of policies (Kalkuhl et al. 2013).

In fact, a recent study by Rochedo et al. (2021) directly imposed low-carbon investments (or increased subsidy rates in associated capital costs) in two IAMs. The Rochedo study showed that stimulus packages of one trillion US$ for the period 2020-2025 (or 200 billion US$/year) will reduce near-term $CO_2$ emissions only to a limited extent compared to the level required for achieving the 1.5 °C target. This finding stands in contrast with the A20's claim that "[a] modest fraction of current global stimulus funds can put the world on track to achieve Paris Agreement goals" based on IAMs simulating a comparable amount of additional low-carbon investments (i.e. 300 billion US$/year in Fig. 1a) but driven effectively by carbon prices. The finding of Rochedo et al. (2021) may indicate the importance of combining public funding with other policy instruments such as carbon pricing to advance the transition broadly. It adds further evidence to the underestimation of required low-carbon investments by A20 if the shift toward decarbonization takes place primarily through low-carbon investments. Nevertheless, further such insights from IAMs directly simulating energy investments, as well as those from other complementary approaches (Guan et al. 2020; Hourcade et al. 2021; Pollitt et al. 2021), are needed to assess the long-term economic and climate implications of COVID-19 public stimulus packages. In particular it may be appropriate for IAMs to model several classes of investments according to their origin, destination, risk level, and/or expected return rate.

## 2.5 Challenges beyond energy investments

Other pillars of the climate strategy need to go hand in hand with energy investments. Further challenges lie in many existing non-financial barriers: short-term public spending should not detract from developing a legal, institutional, and social framework that promotes growing investments in mitigation and adaptation over the long term. A fulfilment of the Paris Agreement goals could further be supported by non-energy related investments in transport (e.g. urban planning) as well as adaptation (Yeo 2019), which were not considered in the A20 estimates. The social cost of the transformation, including associated operation and maintenance costs as well as economy-wide impacts of energy price changes, can be more substantial than the additional total energy investments alone. There are of course substantial social benefits of climate policy in terms of reduced climate damages and a range of co-benefits (Hänsel et al. 2020; Karlsson et al. 2020).

In summary, while the global climate challenge is a long-term problem that will take many decades to solve, a focus on near-term public spending may give a false promise to the public and policymakers. Irrespective of COVID-19 stimulus packages and despite the net zero emission targets by mid-century announced by a growing number of countries, current policies can lead to a large overshoot of the 2 °C warming (IEA 2021a), implying a need for substantial negative emissions to return to that objective (Azar et al. 2013; Tanaka and O'Neill 2018; Johansson 2021; Tanaka et al. 2021). The effort required to reduce the warming below 2 °C after overshoot is not well understood due to uncertainties in carbon cycle and other feedbacks among other reasons (Boucher et al. 2013; Ciais et al. 2013; Steffen et al. 2018; Melnikova et al. 2021). A rocky road is ahead: substantial, broad, and sustained engagements will be needed for achieving the Paris Agreement long-term targets (Grundmann 2016; Hulme 2020), far beyond the short-term emission drop and rebound associated with the COVID-19 pandemic (Liu et al. 2020; Friedlingstein et al. 2021).